\documentclass[a4paper,11pt]{article}
\pdfoutput=1 

\usepackage[DIV10]{typearea}
\usepackage[T1]{fontenc}
\usepackage[bookmarksopen,colorlinks]{hyperref}
\usepackage{graphicx}
\usepackage[latin1]{inputenc}
\usepackage{amssymb}
\usepackage{color}
\usepackage{textcomp}
\usepackage{slashed}
\usepackage{amsmath,bm}
\usepackage{cite}
\usepackage{authblk}
\usepackage{feynmf}
\usepackage{float}

\hypersetup{
   pdfstartview={FitH},
   pdfpagemode={UseOutlines}
}

\begin{document}

\title{\bf {\Large Implications of Bilinear R-Parity Violation on Neutrinos and Lightest Neutralino Decay in Split Supersymmetry}}

\author{Giovanna Cottin%
  \thanks{\texttt{gfc24@hep.phy.cam.ac.uk}}}
\affil{\small{\it Cavendish Laboratory, University of Cambridge, J.J. Thomson Ave, Cambridge CB3 0HE, UK. }}

\author{Marco A. D\'iaz%
  \thanks{\texttt{mad@susy.fis.puc.cl}}}
\affil{\small{\it Instituto de F\'\i sica, Universidad Cat\'olica de Chile, Av. Vicu\~na Mackenna 4860, Santiago, Chile.}}

\author{Sebasti\'an Olivares%
  \thanks{\texttt{saolivap@cern.ch}}}
\affil{\small{\it School of Physics and Astronomy, University of Edinburgh, Edinburgh, UK.}}

\author{Nicol\'as Rojas%
  \thanks{\texttt{nrojas@ific.uv.es}}}
\affil{\small{\it Instituto de F\'isica Corpuscular CSIC/Universitat de Valencia, Parc Cient\'ific,
calle Catedr\'atico Jos\'e Beltr\'an, 2, E-46980 Paterna, Spain.}}

\date{\today}
\maketitle

\begin{abstract}
We discuss neutrino parameters in addition with the effects of a 
Higgs boson of mass $\sim$ 125 GeV in Split Supersymmetry with Bilinear R-Parity Violation. 
This model allows for the explanation of neutrino masses and mixing angles, and has the 
gravitino as Dark Matter candidate. We find constraints on the parameters 
in the neutrino sector of the model by performing a numerical study of the 
parameter space, and by fitting neutrino oscillation observables and the Higgs mass. 
In addition, we study in detail the decay of the lightest neutralino in this model and 
we realize the importance of the exact neutralino/chargino spectrum in 
the computation of its branching ratios.
\end{abstract}

\section{Introduction}

It is an indisputable fact that the ATLAS and CMS Collaborations of the Large Hadron
Collider (LHC) have discovered a new particle \cite{:2012gk,Aad:2014aba,Chatrchyan:2013lba,:2012gu}, with mass near
125 GeV and consistent with the Higgs boson \cite{Englert:1964et,Higgs:1964pj,Higgs:1964ia,Guralnik:1964eu,Higgs:1966ev,Kibble:1967sv} of the Standard Model 
\cite{Glashow:1961tr,Weinberg:1967tq,Salam:1968rm,'tHooft:1972fi}. Its measured value has considerable impact on supersymmetric models, such as the Minimal 
Supersymmetric Standard Model (MSSM). In addition, current LHC searches pushes supertparner masses 
above 1 TeV \cite{Aad:2014wea,Aad:2015mia,Aad:2015gna,Aad:2014qaa,Chatrchyan:2013xna,Chatrchyan:2013mya,Aad:2013wta,Khachatryan:2015vra}. 
This fact leaves the naturalness of the minimal theory in tension and points to an empirically favored 
supersymmetric scenario called Split Supersymmetry (SS) \cite{ArkaniHamed:2004fb,Giudice:2004tc}, where all 
sfermions are very heavy, placed universally at a scale $\tilde{m}$, while charginos and neutralinos remain light. 
Although SS is unnatural by construction and hierarchy is not longer a guiding principle 
(so the Higgs mass has to be fined-tuned), this model retains unification of gauge couplings, 
naturally suppressed flavour mixing and a Dark Matter candidate. For completeness, we mention also the 
alternative scenarios Inverted Hierarchy \cite{Barger:1999iv,Baer:1999md}, High Scale Supersymmetry \cite{Arbey:2011ab}, 
and Intermediate Scale Supersymmetry \cite{Hall:2014vga}.

A very striking effect of Split Supersymmetry is the long lifetime of the
gluino \cite{Gambino:2005eh}. Since all squarks are very heavy, with a mass of order of 
the split supersymmetric scale $\widetilde m$, the gluino will decay via off-shell
squarks, and with an increasing lifetime as $\widetilde m$ increases. Searches have been made 
for long lived gluinos at the LHC with negative results. CMS rules
out gluino R-hadrons with mass $m_{\tilde g}<1$ TeV if their lifetime satisfies
$10^{-6}<\tau_{\tilde g}<10^3$ sec \cite{Khachatryan:2015jha}. ATLAS rules out stable gluinos (gluinos 
that escape the detector before decaying) with mass $m_{\tilde g}\lesssim 1270$ GeV \cite{ATLAS:2014fka}. 
Searches with gluinos decaying fast have been made at ATLAS also with negative results
\cite{Aad:2014wea2,Aad:2015mia, Aad:2015rba}. Analogous searches by CMS give equally negative results, with gluino 
masses bounded from below by 1.26 TeV, unless the LSP has a large mass, in which case the bound
decreases \cite{Chatrchyan:2013iqa}. See also 
\cite{Kilian:2004uj,Hewett:2004nw,Jung:2013zya,Alves:2011ug,Wang:2006gp,Gupta:2004dz,Cheung:2004ad}.

If one allows R-parity to be not conserved, neutrino masses can be generated 
\cite{Dreiner:2010ye,Hirsch:2000ef,Diaz:1997xc,Hempfling:1995wj,deCarlos:1996du}. This can be done 
without introducing problems with too fast proton decay \cite{Nath:2006ut}.
This is so because neutrino masses need only Lepton number violation, while proton decay
needs both Lepton and Baryon number violation. Another issue to be considered is that,
if R-Parity is conserved, the lightest neutralino is a Dark Matter candidate
\cite{Choudhury:2012tc,Baer:2012uya}, but this is no longer the case if R-Parity is violated.
Nevertheless, in models with R-Parity violation the gravitino can be a good dark matter
candidate since it can live longer than the age of the universe \cite{Buchmuller:2007ui,Bailly:2009pe,Restrepo:2011rj}. If one ask the gravitino to be responsible for the positron excess seen by the AMS2 experiment
\cite{Aguilar:2013qda}, then BRpV would not be enough \cite{Carquin:2015uma}. Nevertheless, it is not 
clear that the excess is due to Dark Matter \cite{Hooper:2008kg}, and gravitino as Dark Matter candidate works as 
long as its mass is not larger than ${\cal{O}}(10)$ GeV \cite{Buchmuller:2007ui}.

In this article we study the implications of a Higgs boson mass of $\sim$ 125 GeV on a Split 
Supersymmetric model, which includes R-Parity bilinearly violated terms (SS-BRpV). It has been shown 
possible to accommodate the observed Higgs mass in SS, which imposes constrains in the 
($\tilde{m},\tan\beta$) plane \cite{Giudice:2011cg,ArkaniHamed:2012gw,Arvanitaki:2012ps}. We check that in this case the 
split supersymmetric scale $\widetilde m$ is rather low ($<10^6$ GeV). We also check that the case 
$\tan\beta=1$ is not ruled out, as it is in the MSSM, because of lack of cancellation between 
quark and squark loops. The price to pay may be the divergence of the top quark Yukawa 
coupling at scales larger than $\widetilde m$ but smaller than $M_P$. This does not 
excludes the scenario, but implies the appearance of new physics at that scale. 

We also discuss neutrino masses and mixing angles in SS-BRpV. Neutrino masses arises in this model due to mixing in 
the neutralino/neutrino sector with the inclusion of a gravity induced term \cite{Diaz:2009yz}. 
We discuss how the effect of introducing a constraint on 
the Higgs mass affects the model parameters, requiring that current
experimental values from neutrino physics given in \cite{Tortola:2012te,Forero:2014bxa} 
are reproduced with a 95\% confidence level. In particular, we see the model forces a strong dependence on 
the atmospheric and solar neutrino mixing angles. In addition, we study in 
detail the two-body decays of the lightest neutralino. We conduct a general scan of our available parameter space 
and realize the importance in knowing the exact neutralino/chargino spectrum in the 
computation of the neutralino branching fractions.

\section{Split Supersymmetry and the Higgs Mass}

The split supersymmetric lagrangian below the $\widetilde m$ scale includes
charginos, neutralinos, plus all the SM particles, including the SM-like Higgs boson
$H$ \cite{ArkaniHamed:2004fb,Giudice:2004tc}. The lagrangian looks as follows,
\begin{eqnarray}
{\cal L}^{split}_{susy}&=& {\cal L}^{split}_{kinetic} \ + \ m^2H^\dagger 
H - \frac{\lambda}{2}(H^\dagger H)^2 
-\Big[ Y_u \overline q_L u_R i \sigma_2 H^* \ + \ Y_d \overline q_L d_R H \ 
+ \ Y_e \overline l_L e_R H \ + \
\nonumber\\ 
&&
+ \frac{M_3}{2} \widetilde G\widetilde G \ + \ \frac{M_2}{2} \widetilde W 
\widetilde 
W \ + \ \frac{M_1}{2} \widetilde B \widetilde B \ + \ 
\mu \widetilde H_u^T i \sigma_2 \widetilde H_d \ + \
\label{LagSplit}\\ &&
+\textstyle{\frac{1}{\sqrt{2}}} H^\dagger
(\tilde g_u \sigma\widetilde W \ + \ \tilde g'_u\widetilde B)\widetilde H_u
 \ + \ \textstyle{\frac{1}{\sqrt{2}}} H^T i \sigma_2
(-\tilde g_d \sigma \widetilde W+ \tilde g'_d \widetilde B)\widetilde H_d 
+\mathrm{h.c.}\Big],
\nonumber
\end{eqnarray}
In the gaugino sector we use as input the low energy values for the Bino and Wino 
masses $M_1$ and $M_2$, and the higgsino mass $\mu$. At a scale $M_\chi$ we 
decouple the gauginos and higgsinos, such that below that scale the SM is valid.
To calculate the Higgs mass in this model we first need the RGE evolution of the quartic 
Higgs coupling, and second the quantum corrections, that we approximate at one loop following
a prescription for the renormalization scale given in ref.~\cite{Bernal:2007uv}.

In Split Supersymmetry a unification of gauge couplings is assumed \cite{ArkaniHamed:2004fb}.
We start at the electroweak scale $m_Z$ with 
SM-RGE, changing at the scale $M_\chi$ to SS-RGE, and changing again at the 
$\widetilde m$ scale to the MSSM-RGE \cite{Giudice:2004tc}. 
The initial condition is given by the 
values of the gauge couplings $g_1$, $g_2$, and $g_3$, at the weak scale. We 
calculate the electroweak couplings with the help of 
$\alpha_{\text{fin}}^{-1}(m_Z)=128.962\pm0.014$ \cite{Hoecker:2010qn}, and 
$\sin^2\theta_w(m_Z)=0.23119\pm0.00014$ \cite{Nakamura:2010zzi}, namely 
$g_2^2=4\pi\alpha_{fin}/s^2_w$ and $g_1^2=5g'^2/3$, $g'^2=4\pi\alpha_{fin}/c^2_w$. 
In turn, the strong coupling constant satisfy $g_3^2=4\pi\alpha_s$, with
$\alpha_s(m_Z)=0.1184\pm0.0007$ \cite{Nakamura:2010zzi}. The intersection of 
the three gauge coupling RGE curves defines the Grand Unification scale $M_{GUT}$.
Since the unification is not perfect (within experimental errors), we define $M_{GUT}$ 
as the average of the three meeting points. 

Matching conditions at the scale $\widetilde m$ between SS and the MSSM are,
\begin{eqnarray}
\tilde g_u(\tilde m)=g(\tilde m) \ \sin\beta
&,\qquad&
\tilde g_d(\tilde m)=g(\tilde m) \ \cos\beta
\nonumber\\
\tilde g'_u(\tilde m)=g'(\tilde m) \ \sin\beta
&, \qquad &
\tilde g'_d(\tilde m)=g'(\tilde m) \ \cos\beta,
\label{gtildeBC}
\end{eqnarray}
The large difference that may appear between up and down $\tilde g$ couplings at $\widetilde m$
is due to the value of $\tan\beta$.

%
\begin{figure}[ht]
\centering
\includegraphics[width=0.6\textwidth,angle=0]{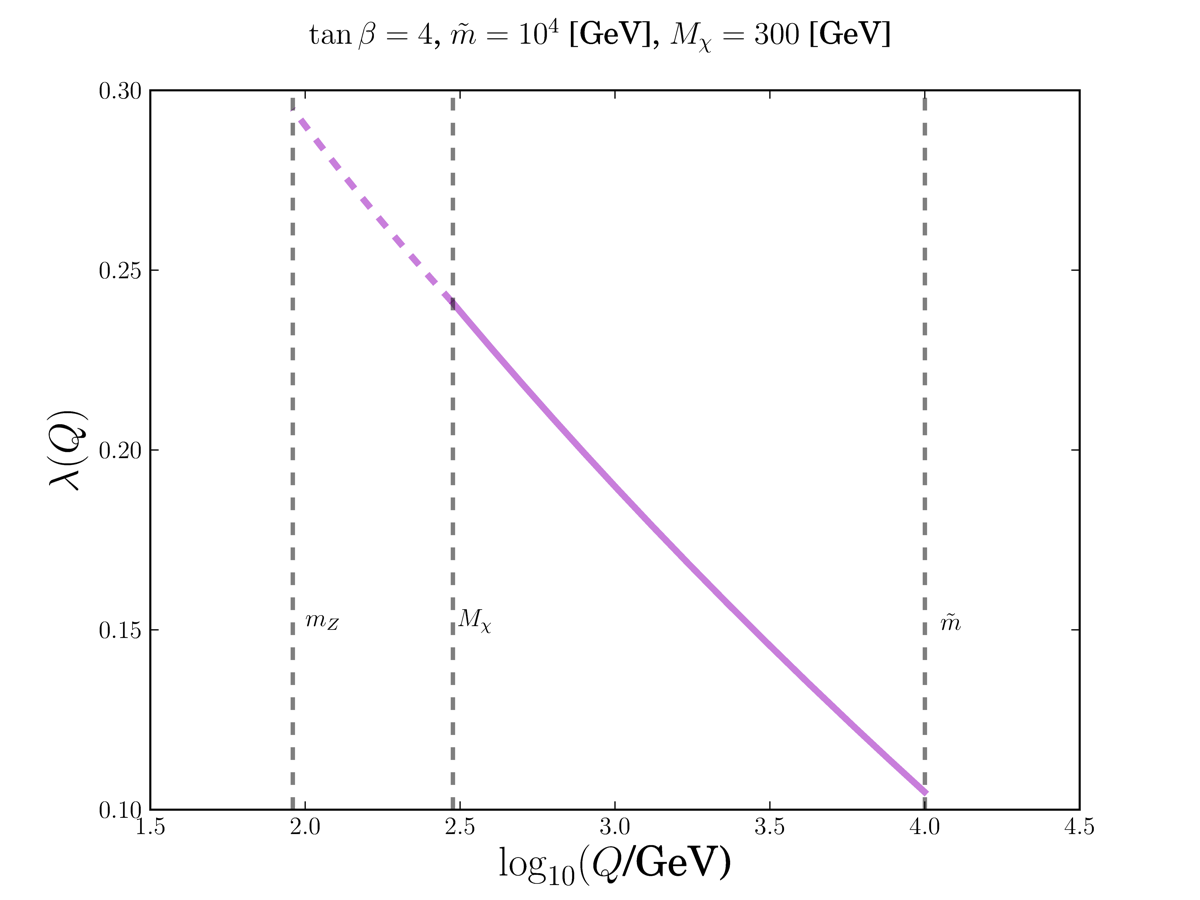}
\caption{ Running of the Higgs coupling $\lambda$, for $\tan\beta=4$,
$\widetilde m=10^{4}$ GeV and $M_\chi=300$ GeV.
}
\label{LAMBDAT4M4}
\end{figure}
%
In fig.~\ref{LAMBDAT4M4} we see the running of the Higgs quartic coupling $\lambda$
for the set of input parameters $\tan\beta=4$, $\widetilde m=10^{4}$ GeV,
and $M_{\chi}=300$ GeV. The starting point is also at $\widetilde m$ with the
matching condition,
\begin{equation}
\lambda(\widetilde m) = \frac{1}{4} \left[
g^2(\widetilde m) + g'^2(\widetilde m) \right] \cos^2 2\beta
\end{equation}
As we can see, the threshold at $M_\chi$ has just a small effect. The renormalized Higgs
mass includes the tree-level contribution proportional to the quartic coupling $\lambda$
evaluated at the chosen renormalization scale $Q=m_t$, following ref.~\cite{Bernal:2007uv}.
The value of the Higgs coupling at the renormalization scale is $\lambda=0.264$,
leading to a Higgs boson mass $m_H=125.3$ GeV, consistent with observations from the LHC.

%
\begin{figure}[ht]
\centering
\includegraphics[width=0.6\textwidth,angle=0]{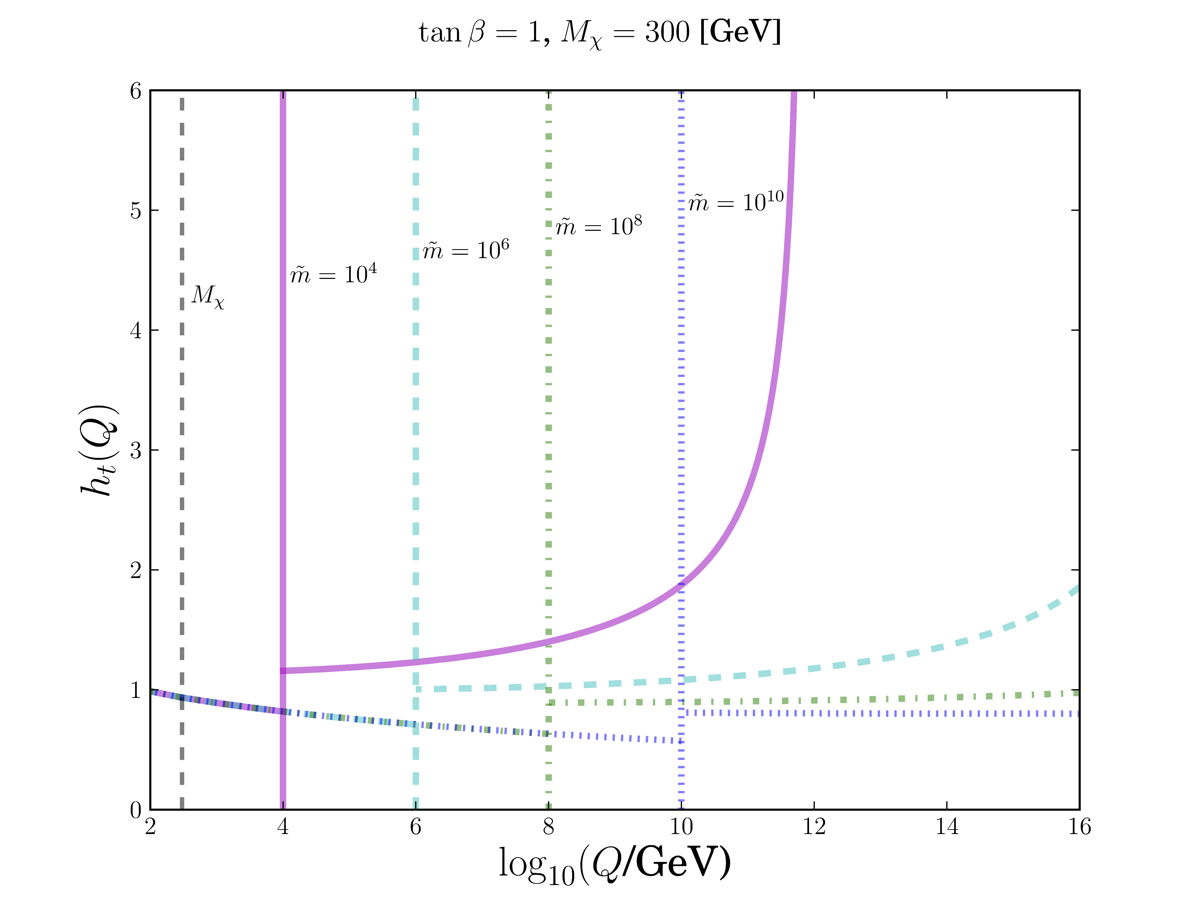}
\caption{ Running of the top quark Yukawa coupling $h_t$ for $\tan\beta=1$
and several values of the Split Supersymmetric scale $\widetilde m$.
}
\label{HT}
\end{figure}
%
For extreme values like $\tan\beta=1$, we can have a value for the Higgs mass consistent 
with the experimental evidence, nevertheless, unification of gauge couplings fails because
the top quark Yukawa coupling becomes non-perturbatively large at a scale larger than
$\widetilde m$. This fact can be seen in fig.~\ref{HT} for different values of the SS scale.
The fact that the top Yukawa coupling diverges is an indication of new physics 
appearing at that scale. The model ceases to be valid beyond that scale. In the figure we show
also the threshold at $\widetilde m$. Below it, the SS-RGE controls the behavior of the top 
quark Yukawa $h_t$ and its evolution is the same for any of the chosen values for $\widetilde m$. 
After that threshold we switch to the MSSM-RGE for $h_t$ which hold the following boundary
condition,
\begin{equation}
h_t^{SS}(\widetilde m) = h_t^{MSSM}(\widetilde m) \cos\beta
\end{equation}
and this explains the discontinuity for $h_t$ at the threshold. We stress the fact that 
the divergence for $h_t$ at a scale larger than $\widetilde m$ does not invalidates the
low scale SS model.

A Higgs mass compatible with experiments is
obtained for a SS scale $10^{4}\lesssim \widetilde m \lesssim 10^{6}$ GeV, and 
any value of $\tan\beta$ is possible (a SS model with $\widetilde m$ smaller than
$10^{4}$ is not much different to the MSSM). The fact that $\tan\beta=1$ with 
$\widetilde m\sim10^{6}$ is consistent with the experimental measurements for the 
Higgs mass is an interesting fact, although already noticed in the literature 
\cite{Giudice:2004tc}. The price we pay in this case is that the top quark Yukawa coupling
becomes non-perturbative at scales larger than $\widetilde m$ and as a consequence
the gauge coupling unification is lost.

%
\begin{figure}[ht]
\centering
\includegraphics[width=0.6\textwidth,angle=0]{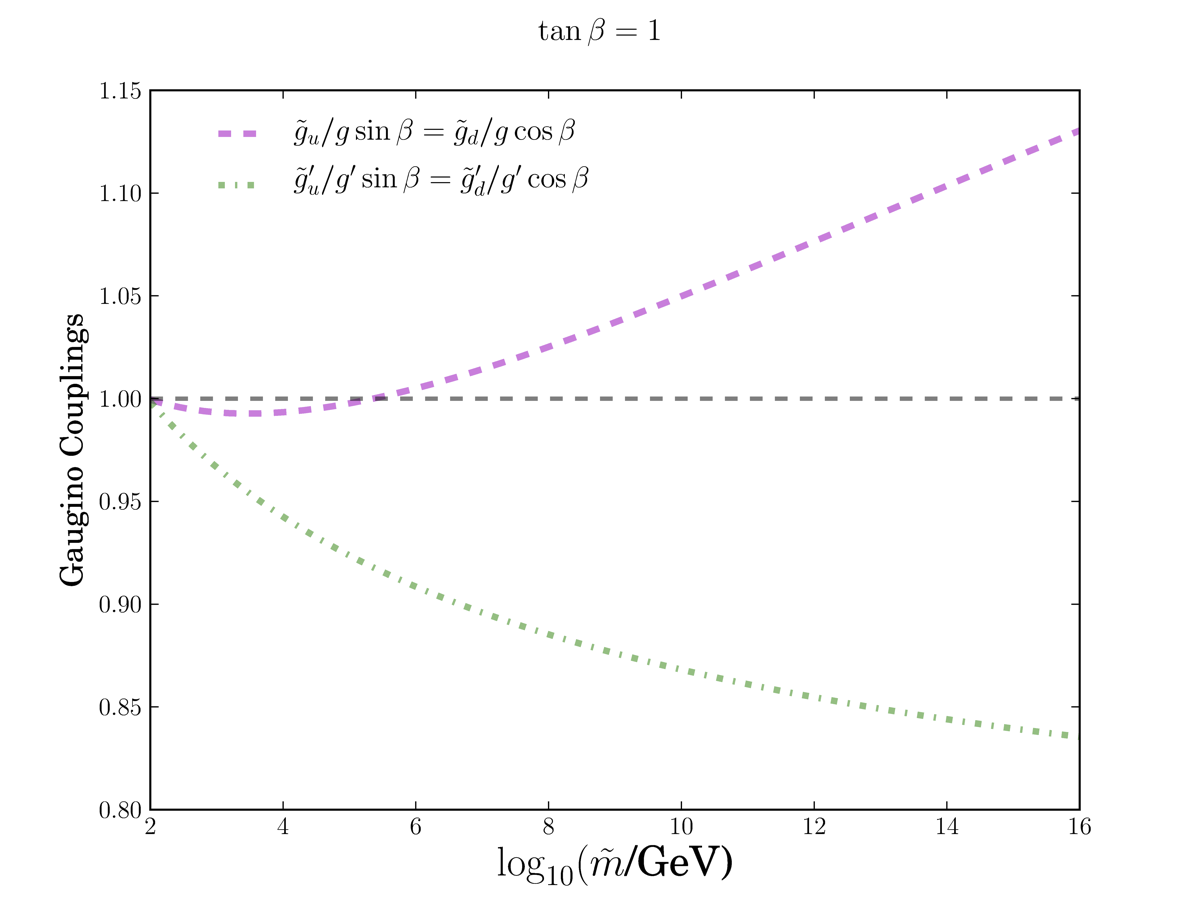}
\caption{ Ratio between the gaugino couplings and the gauge couplings, weighted by
$\sin\beta$ or $\cos\beta$, as a function of the SS scale $\widetilde m$, calculated at the
weak scale for $\tan\beta=1$.
}
\label{GTAN1}
\end{figure}
%
In fig.~\ref{GTAN1} we have the Higgs-higgsino-gaugino 
couplings for the special case $\tan\beta=1$.
From eq.~(\ref{gtildeBC}) we see that in this case both couplings $\tilde g$ and both
$\tilde g'$ are equal to each other at $\widetilde m$, and since RGE are also the 
same, the couplings remain equal, as can be seen in the figure. From the values
$\widetilde m\lesssim 10^6$ GeV we also expect in this case deviations of at most $10\%$.

\section{Neutrino Masses in Bilinear R-Parity Violation}

If R-Parity is bilinearly violated, very little of the above conclusions
are changed, since the RGE are the same. In SS-BRpV, the decoupling of the sleptons
induce BRpV couplings between gauginos, higgsinos and Higgs, which 
at lower scales look like \cite{Diaz:2006ee},
\begin{equation}
{\cal L}^{split}_{RpV}=
\epsilon_i \widetilde H_u^T i \sigma_2 L_i \ -\ 
\textstyle{\frac{1}{\sqrt{2}}} a_i H^T i \sigma_2
(-\tilde g_d \sigma\widetilde W+\tilde g'_d\widetilde B)L_i \ + \ h.c., 
\label{LSplitRpV}
\end{equation}
where $a_i$ are dimensionless parameters that characterize the decoupling of the sleptons.
These terms induce a neutralino/neutrino mixing when the Higgs field acquire 
a vacuum expectation value,
\begin{equation}
{\cal L}^{split}_{RpV} = -\left[
\epsilon_i \widetilde H_u^0 + \frac{1}{2} a_i v \left( 
\tilde g_d \widetilde W_3 - \tilde g'_d \widetilde B \right)
\right] \nu_i \ + \ h.c. \ + \ \ldots
\end{equation}
where $v$ is normalized such that the $W$ gauge boson has a mass $m_W=\frac{1}{2}gv$,
thus $v\approx 246$ GeV. In this way, the neutralino/neutrino sector in the basis
$\psi=(-i\lambda',-i\lambda^3,\widetilde{H}_d^0,
\widetilde{H}_u^0, \nu_{e},\nu_{\mu}, \nu_{\tau} )$
develops a mass matrix that we write as follows,
\begin{equation}
{\cal M}_N^{SS}=\left[\begin{array}{cc} {\mathrm M}_{\chi^0}^{SS} & (m^{SS})^T \\ 
m^{SS} & 0 \end{array}\right],
\label{X07x7}
\end{equation}
where ${\mathrm M}_{\chi^0}^{SS}$ is the neutralino mass sub-matrix,
\begin{equation}
{\bf M}_{\chi^0}^{SS}=\left[\begin{array}{cccc}
M_1 & 0 & -\frac{1}{2}\tilde g'_d v & \frac{1}{2}\tilde g'_u v \\
0 & M_2 & \frac{1}{2}\tilde g_d v & -\frac{1}{2}\tilde g_u v \\
-\frac{1}{2}\tilde g'_d v & \frac{1}{2}\tilde g_d v & 0 & -\mu \\
\frac{1}{2}\tilde g'_u v & -\frac{1}{2}\tilde g_u v & -\mu & 0
\end{array}\right],
\label{X0massmat}
\end{equation}
and $m^{SS}$ includes the mixing between neutralinos and neutrinos,
\begin{equation}
m^{SS}=\left[\begin{array}{cccc}
-\frac{1}{2} \tilde g'_d a_1v & \frac{1}{2} \tilde g_d a_1v 
& 0 &\epsilon_1 \cr
-\frac{1}{2} \tilde g'_d a_2v & \frac{1}{2} \tilde g_d a_2v&0 
& \epsilon_2 \cr
-\frac{1}{2} \tilde g'_d a_3v & \frac{1}{2} \tilde g_d a_3v&0 
& \epsilon_3
\end{array}\right].
\end{equation}

The mass matrix in eq.~(\ref{X07x7}) can be block-diagonalized, and an effective 
neutrino $3\times3$ mass matrix is generated,
\begin{equation}
{\bf M}_\nu^{eff}=
\frac{v^2}{4\det{M_{\chi^0}}}
\left(M_1 \tilde g_d^2 + M_2 \tilde g'^2_d \right)
\left[\begin{array}{cccc}
\lambda_1^2        & \lambda_1\lambda_2 & \lambda_1\lambda_3 \cr
\lambda_2\lambda_1 & \lambda_2^2        & \lambda_2\lambda_3 \cr
\lambda_3\lambda_1 & \lambda_3\lambda_2 & \lambda_3^2
\end{array}\right],
\label{treenumass}
\end{equation}
where the determinant of the neutralino mass matrix is:
\begin{equation}
\det{M_{\chi^0}}=-\mu^2 M_1 M_2 + \frac{1}{2} v^2\mu \left( 
M_1 \tilde g_u \tilde g_d + M_2 \tilde g'_u \tilde g'_d \right)
+\textstyle{\frac{1}{16}} v^4 
\left(\tilde g'_u \tilde g_d - \tilde g_u \tilde g'_d \right)^2 .
\label{detNeut}
\end{equation}
The $\lambda_i$ parameters in eq.~(\ref{treenumass}) are defined as
$\lambda_i\equiv a_i\mu+\epsilon_i$. 

We follow the model explained in ref.~\cite{Diaz:2009yz}, where the solar
neutrino mass is generated by a non-renormalizable dimension 5 operator
generated by an unknown quantum gravity theory. The strength of this operator
is characterized by the parameter $\mu_g$, which has dimensions of mass. 
Alternative scenarios are Partial Split Supersymmetry \cite{Diaz:2006ee}, where the $\mu_g$
term is generated by uncanceled contributions from Higgs bosons, and SUSY
models with Trilinear $R_p$ violation \cite{Chun:2003xf}, where the $\mu_g$ term can be generated by 
the trilinear couplings.
In this context, the generated neutrino mass matrix is,
\begin{equation}
\label{eq_mixmod}
M_\nu^{ij}=A \lambda^i\lambda^j+\mu_g
\end{equation}
where $A$ can be read from eq.~(\ref{treenumass}). In this case, one of the
neutrinos remain massless, and the other two acquire the following mass,
\begin{equation}
m_{\nu_{2,3}} = \frac{1}{2}\left( A|\vec\lambda|^2+
3\mu_g\right) \pm \frac{1}{2}\sqrt{
\left( A|\vec\lambda|^2+3\mu_g\right)^2-
4A\mu_g|\vec v\times\vec\lambda|^2}
\end{equation}
where we have used the auxiliary vector $\vec v=(1,1,1)$. In ref.~\cite{Diaz:2009yz}
it was proved that the experimental results on neutrino physics force
$\mu_g\approx 3\times 10^{-3}$ eV. If we also have $\mu_g\ll A|\vec\lambda|^2$, the
atmospheric and solar mass squared are,
\begin{eqnarray}\label{eq_msolapprox}
\Delta m^2_{atm} &=& A^2 \vec{\lambda}^4+2 A \mu_g
(\vec{v}\cdot\vec{\lambda})^2+{\mathcal{O}}(\mu_g^3) \nonumber\\
\Delta m^2_{sol} &=& \mu_g^2\frac{(\vec{v}\times
\vec{\lambda})^4}{\vec{\lambda}^4}+{\mathcal{O}}(\mu_g^3)
\label{eq_matmapprox}
\end{eqnarray}
%
%
\begin{figure}[ht]
\centering
\includegraphics[width=0.45\textwidth,angle=0]{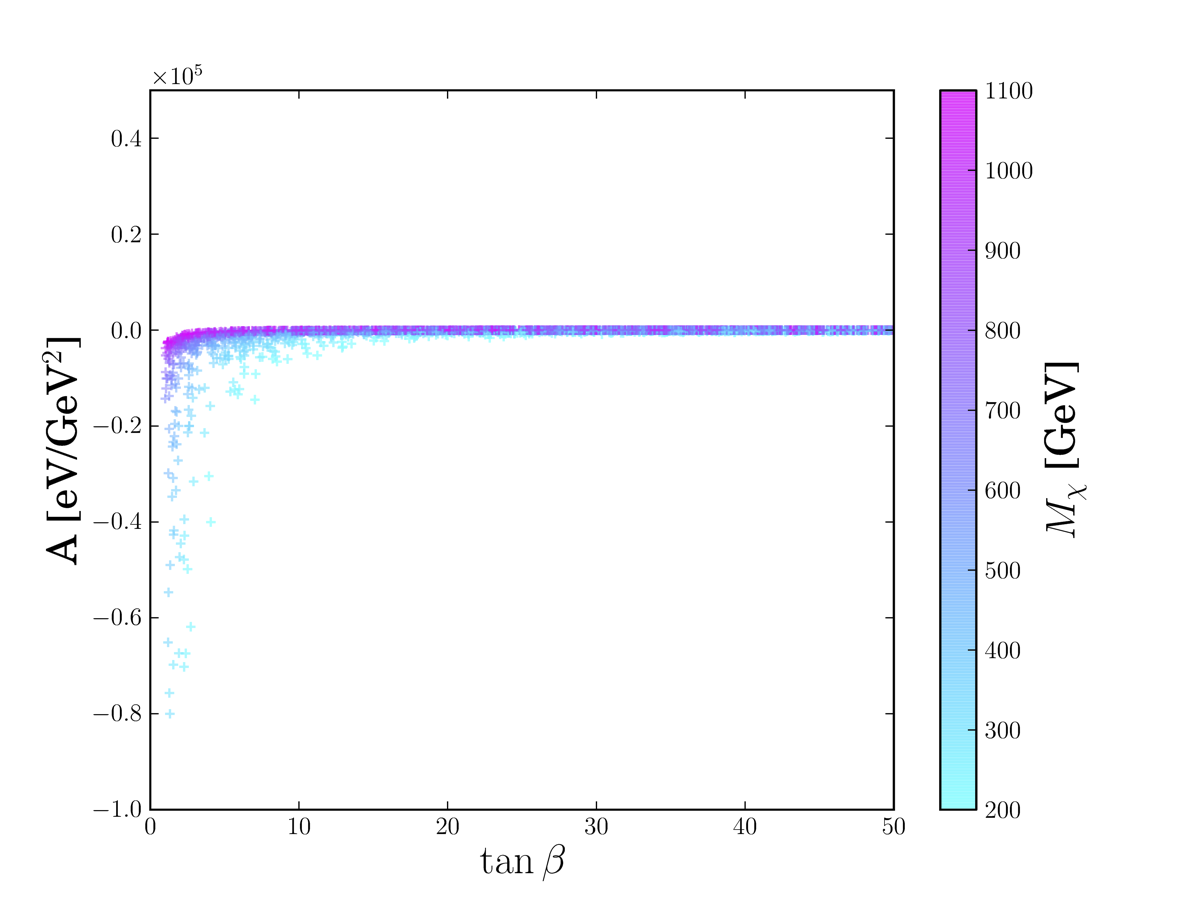}
\includegraphics[width=0.45\textwidth,angle=0]{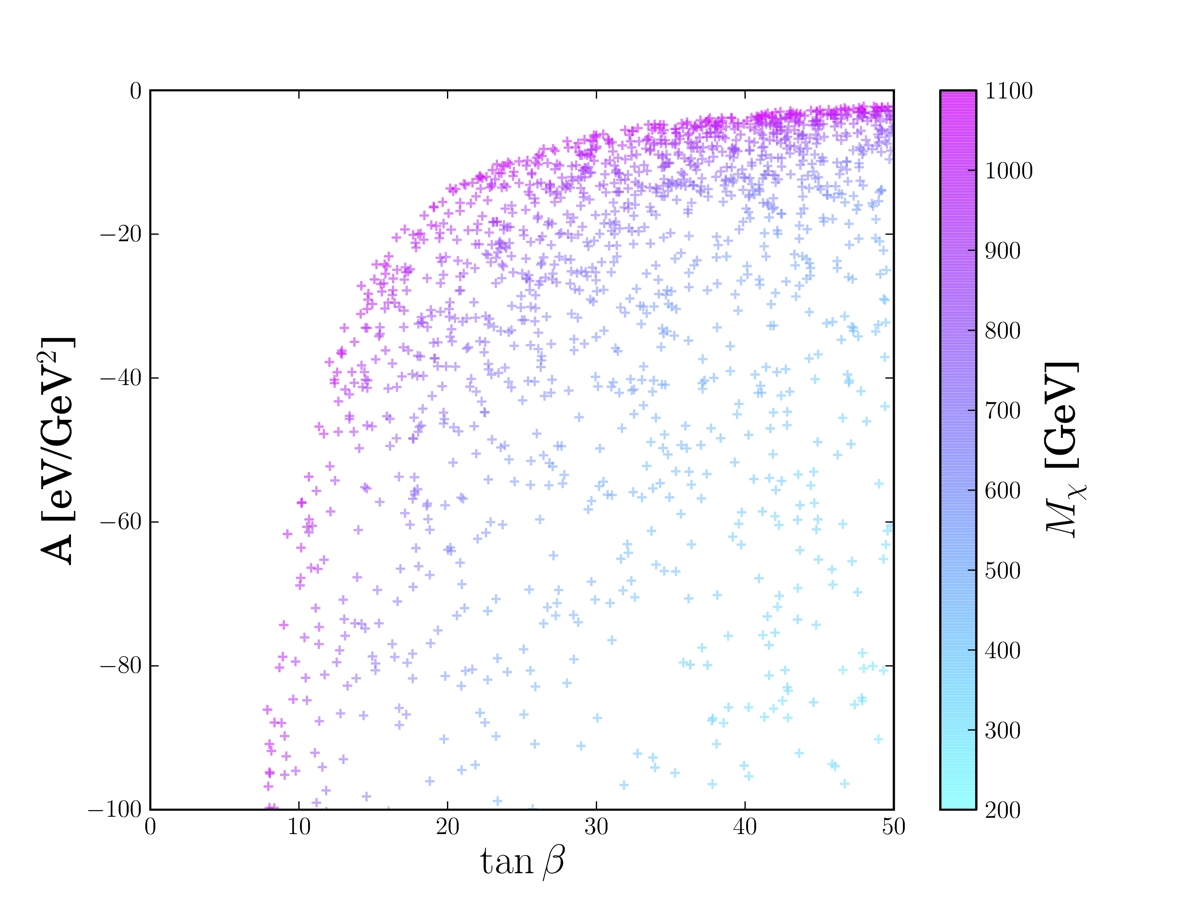}
\caption{ Coefficient $A$ as a function of $\tan\beta$ for different values
of the $M_\chi$ scale, with a blow-up for lower values of $|A|$ in the right frame.}
\label{A}
\end{figure}
%
The value of $A$ can be directly calculated from the R-Parity conserving parameters
we have been working with in the previous sections: $\tan\beta$, $\widetilde m$, and
$M_\chi$, but with the addition that $\widetilde m$ is determined as a function of $\tan\beta$
such that we get a Higgs boson mass according to the experimental observation.

The result can be seen in fig.~\ref{A}, 
where we have the value of $A$ as a function of $\tan\beta$ for different values 
of $M_\chi$. Both signs for $A$ are possible, obtained by switching the sign of the gaugino mass 
parameters. Absolute values of $A$ can be of several thousands for small values of $\tan\beta$
as well as a few units ($\mathrm{eV/GeV}^2$) for large $\tan\beta$ and large $M_\chi$.
It is a characteristic of this model that the value of the Higgs mass measured at the LHC forces
large values of $|A|$.
%
\begin{figure}[ht]
\centering
\includegraphics[width=0.45\textwidth,angle=0]{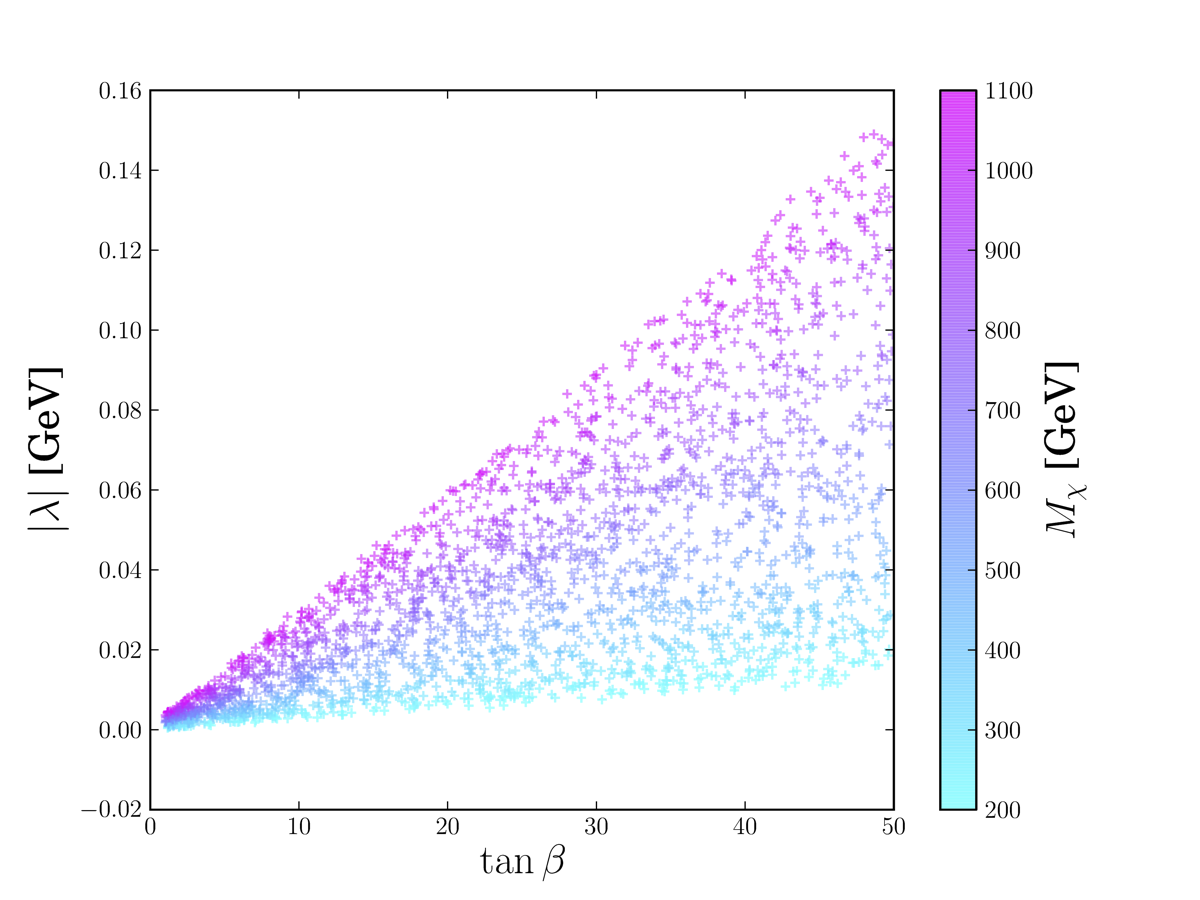}
\includegraphics[width=0.45\textwidth,angle=0]{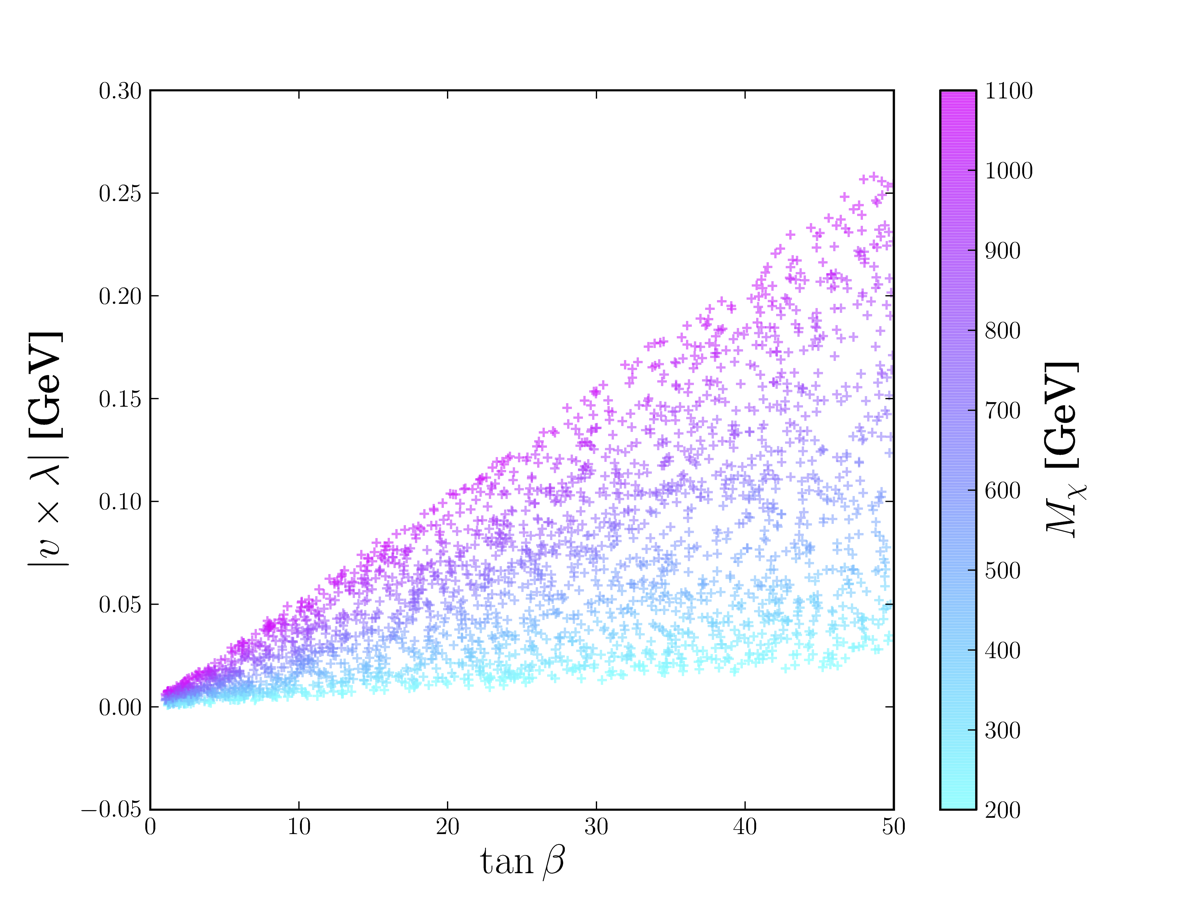}
\caption{ Values for $|\vec\lambda|$ (left) and $|\vec{v}\times\vec{\lambda}|$
(right) as a function of $\tan\beta$ for several different values of $M_\chi$.
}
\label{LAMBDA}
\end{figure}
%
According to eq.~(\ref{eq_matmapprox}) the atmospheric mass squared difference, which 
experimentally is $\Delta m^2_{atm} \approx 2\times10^{-3} \, {\mathrm{eV}^2}$, is to
first order equal to $A^2\vec{\lambda}^4$, and this allow us to calculate $|\vec\lambda|$
in each of the scenarios defined by $\tan\beta$ and $M_\chi$. In fig.\ref{LAMBDA}-right
we have $|\vec\lambda|$ as a function of $\tan\beta$ for different values of $M_\chi$.
It increases with $\tan\beta$ and with $M_\chi$ because $A$ does the opposite. Similarly
according to eq.~(\ref{eq_matmapprox}) the solar mass squared difference, which 
experimentally satisfies $\Delta m^2_{sol} \approx 8\times10^{-5} \, {\mathrm{eV}^2}$,
is up to first order equal to 
$\mu_g^2{(\vec{v}\times\vec{\lambda})^4}/{\vec{\lambda}^4}$, thus we can determine
$|\vec{v}\times\vec{\lambda}|$ in each of the scenarios.
We see the result in fig.\ref{LAMBDA}-left, with a similar result compared 
to $|\vec\lambda|$, just typically twice as large. 

\begin{figure}[ht]
\centering
\includegraphics[width=0.45\textwidth,angle=0]{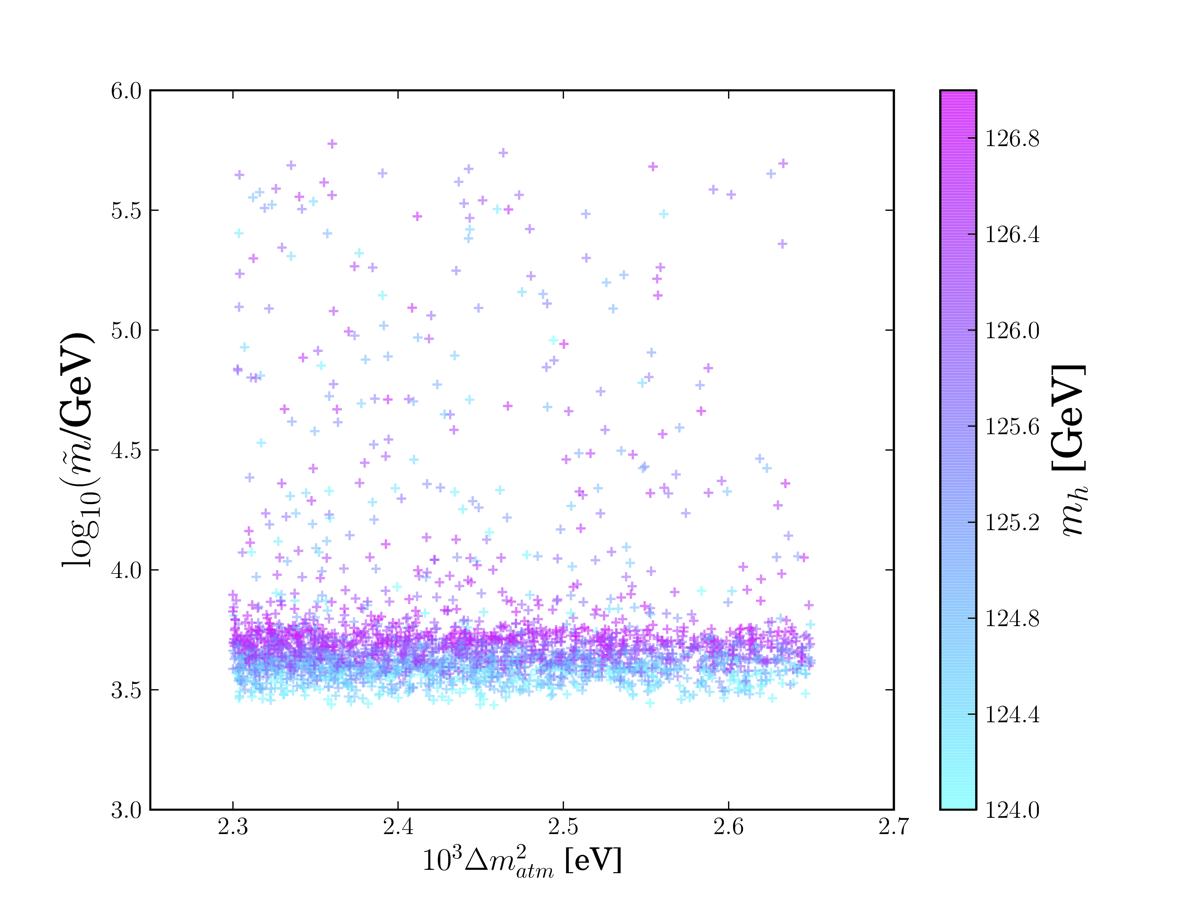}
\includegraphics[width=0.45\textwidth,angle=0]{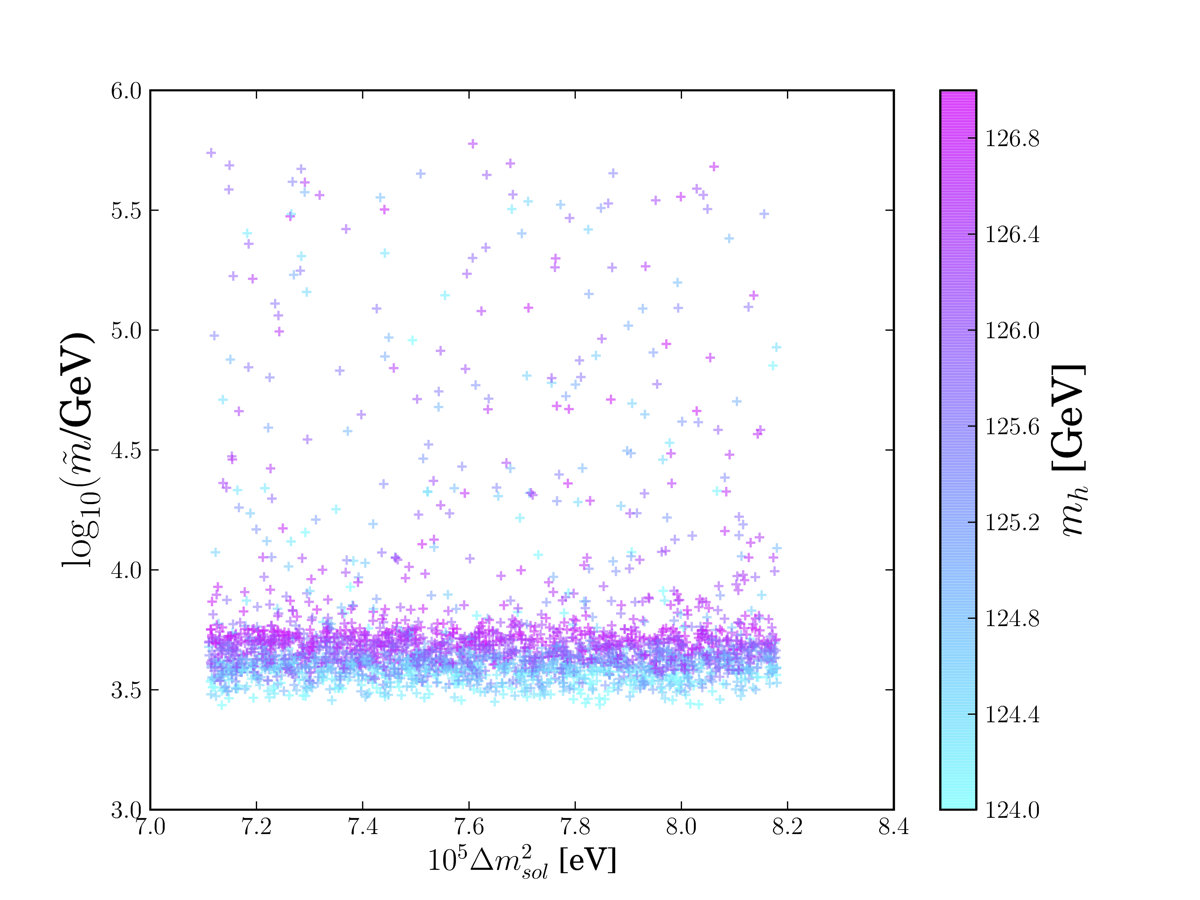}
\caption{ Values for $\Delta m^{2}_{atm}$ (left) and $\Delta m^{2}_{sol}$
(right) as a function of $\tilde{m}$ for different values of the allowed Higgs mass.
}
\label{Deltam}
\end{figure}

\begin{figure}[ht]
\centering
\includegraphics[width=0.45\textwidth,angle=0]{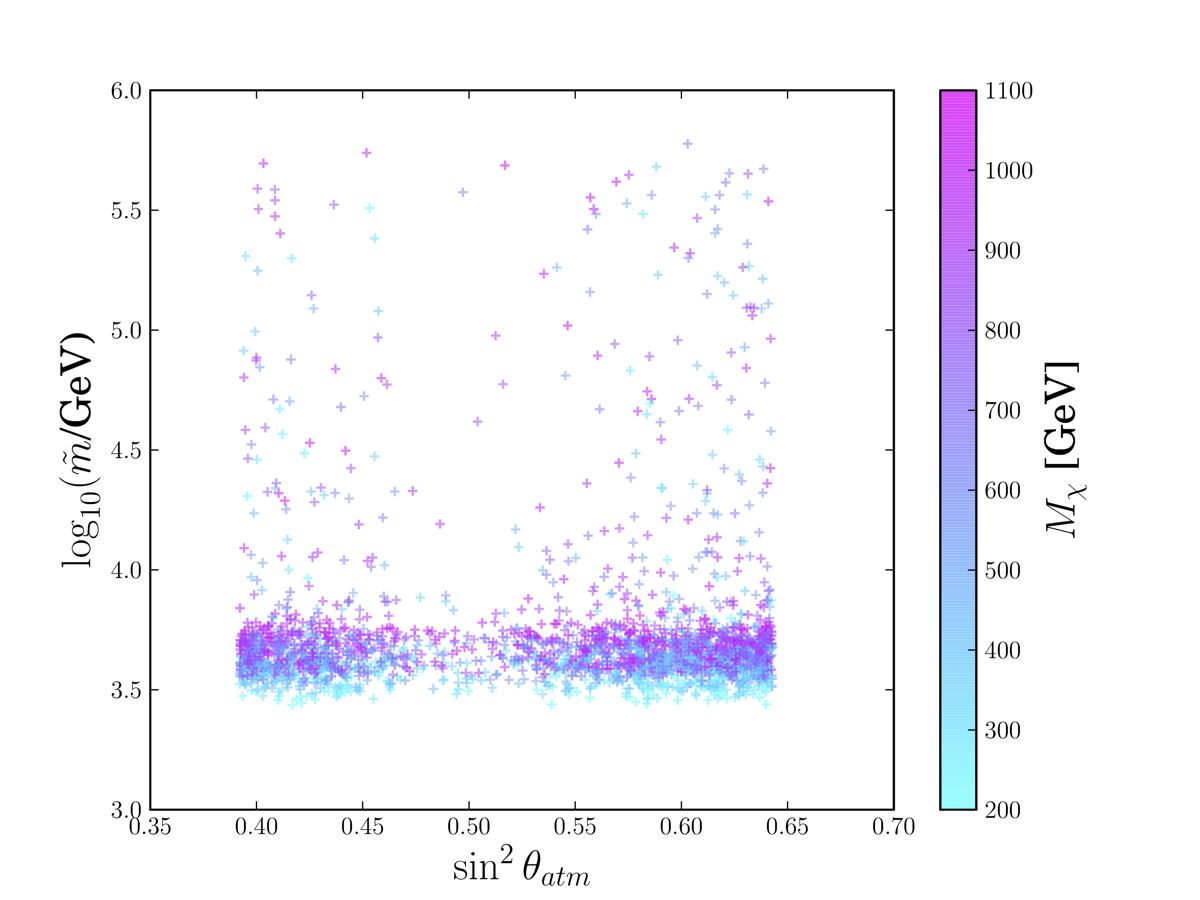}
\includegraphics[width=0.45\textwidth,angle=0]{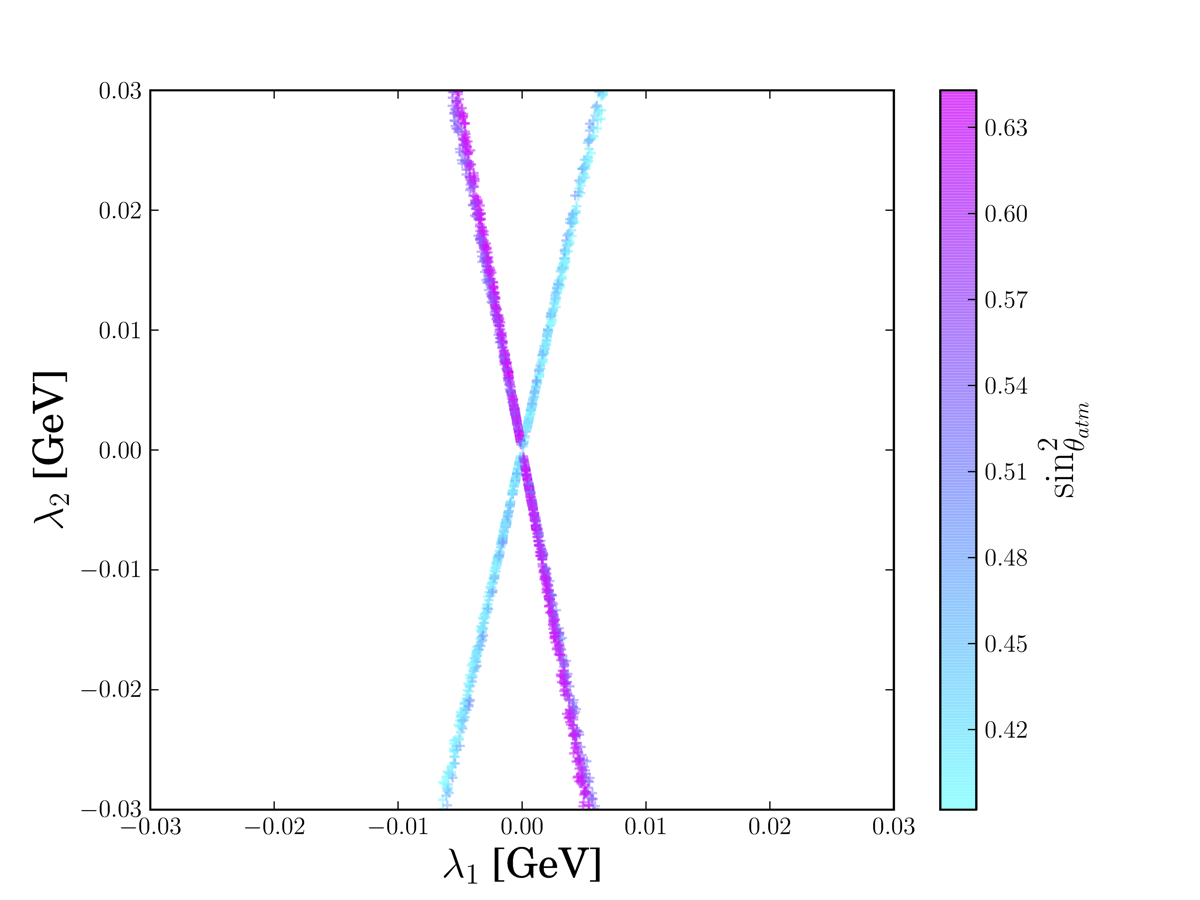}
\caption{ Values for $\sin^{2}{\theta}_{atm}$ as a function of $\tilde{m}$ for for several different values of $M_\chi$ (left). We also show the dependence of $\sin^{2}{\theta}_{atm}$ with the BRpV parameters $\lambda_{1}$ and $\lambda_{2}$ (right).
}
\label{sin2solatm}
\end{figure}

\begin{figure}[H]
\centering
\includegraphics[width=0.45\textwidth,angle=0]{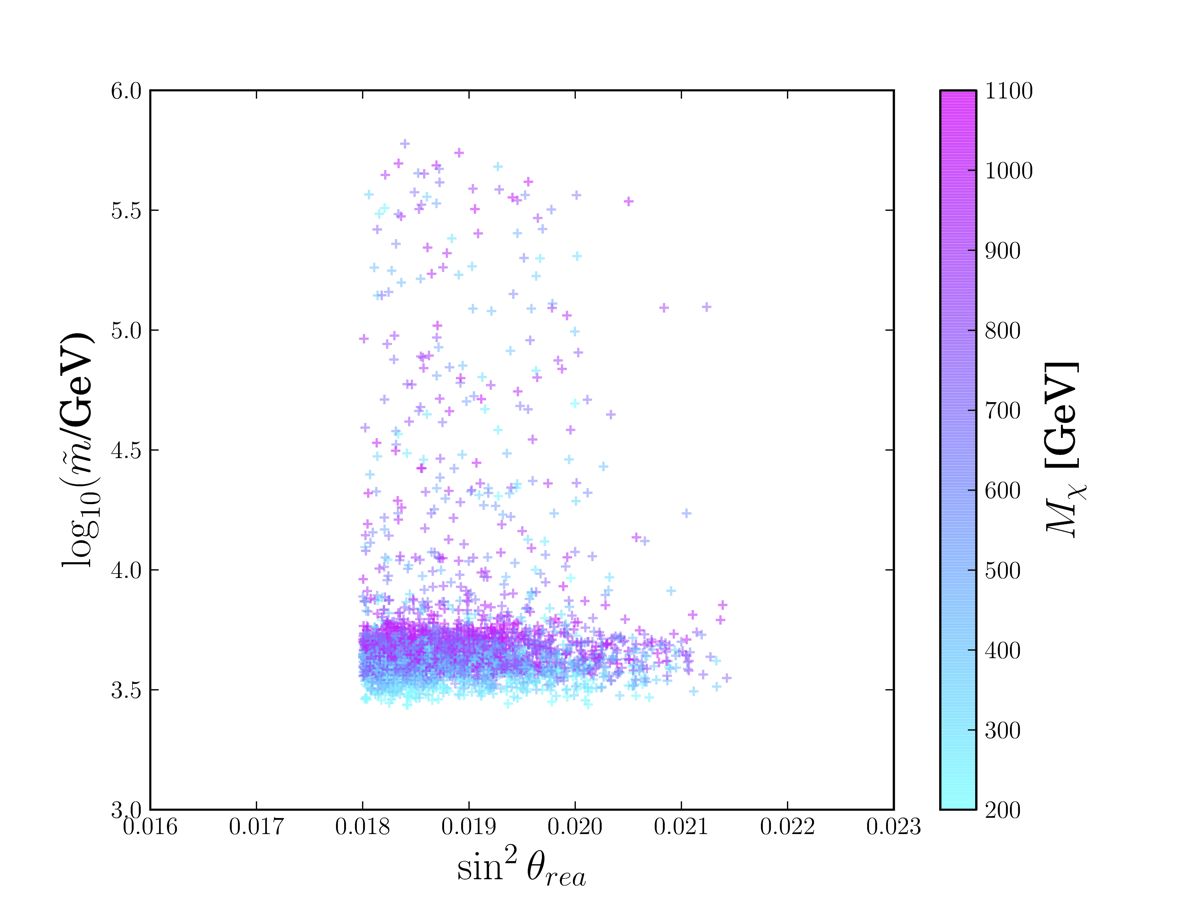}
\includegraphics[width=0.45\textwidth,angle=0]{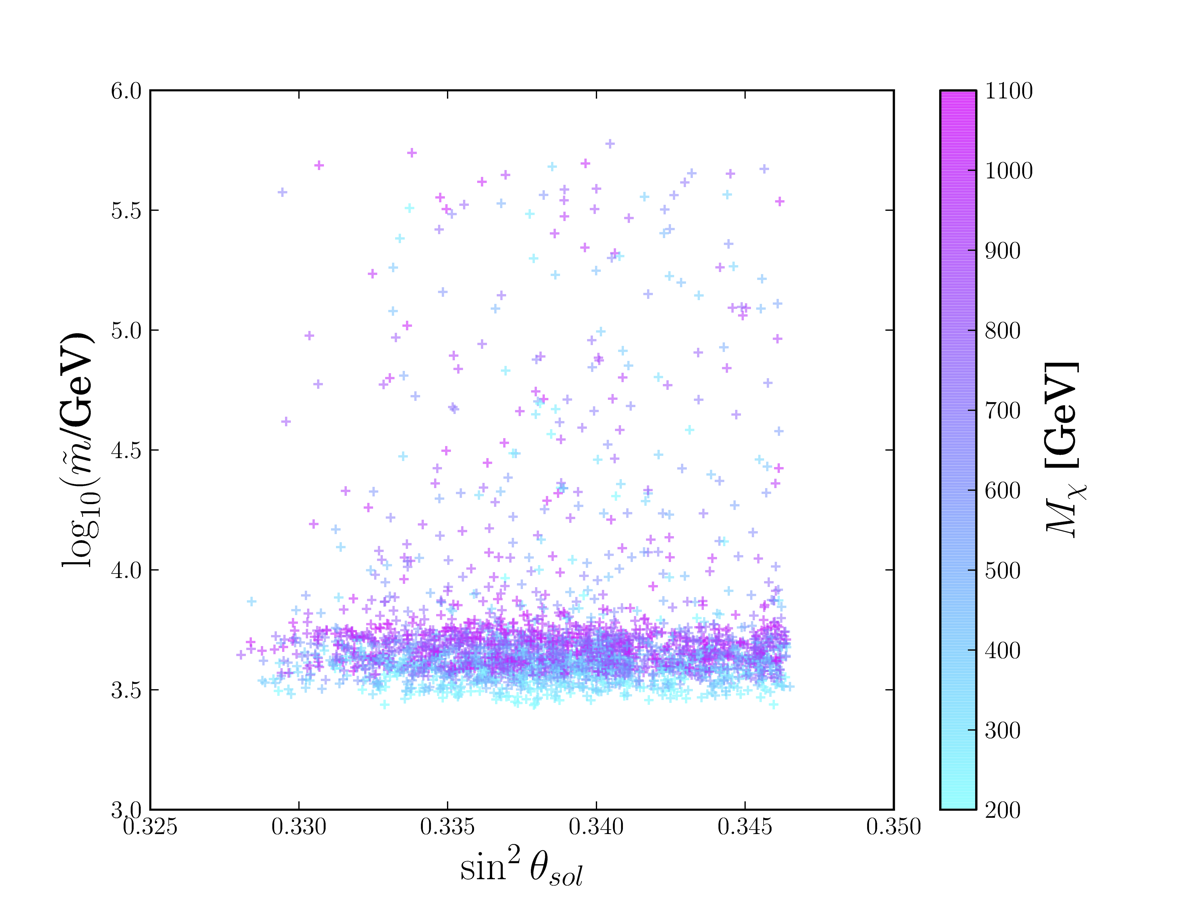}
\caption{ Values for $\sin^{2}{\theta}_{reac}$ (left) and $\sin^{2}{\theta}_{sol}$ (right)  as a function of $\tilde{m}$  for several different values of $M_\chi$.
}
\label{sin2reac_sol2atm2}
\end{figure}

\begin{figure}[ht]
\centering
\includegraphics[width=0.7\textwidth,angle=0]{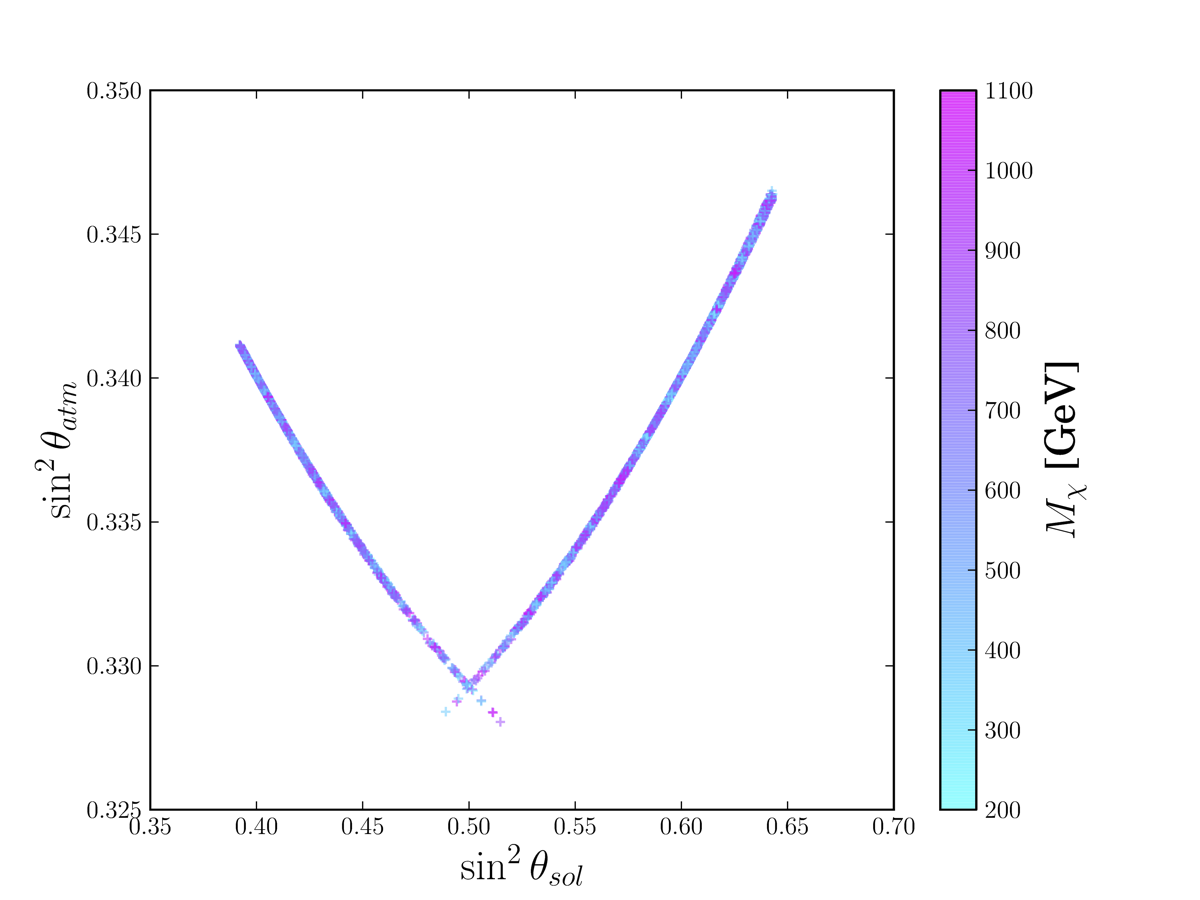}
\caption{ Allowed region in the $\sin^{2}{\theta}_{sol}$-$\sin^{2}{\theta}_{atm}$ plane for several different values of $M_\chi$.
}
\label{sin2atmsin2sol}
\end{figure}

Now we can compute the atmospheric and solar mass squared differences, which can be seen in fig. \ref{Deltam}. 
We notice that, independently of the allowed value of the mass squared differences, the Higgs mass grows with $\tilde{m}$. 
%
%
Neutrino mixing angles depend on the values of all the $\lambda_i$. For a diagonal charged lepton matrix, 
these are given by \cite{Diaz:2009yz}
\begin{equation}
\sin^{2}{\theta_{reac}}=\frac{\lambda^{2}_{1}}{|\vec{\lambda}|^{2}} \hspace{0.2cm},\hspace{0.2cm}
\tan^{2}{\theta_{atm}}=\frac{\lambda^{2}_{2}}{\lambda^{2}_{1}} \hspace{0.2cm},\hspace{0.2cm}
\tan^{2}{\theta_{sol}}=\frac{(\lambda^{2}_{2}-\lambda_{1}\lambda_{2}-\lambda_{1}\lambda_{3}+\lambda^{2}_{3})^{2}}{(\lambda^{2}_{1}+\lambda^{2}_{2}+\lambda^{2}_{3})(\lambda_{3}-\lambda_{2})^{2}} .
\end{equation}
On fig. \ref{sin2solatm} left we see atmospheric angle against $\tilde{m}$ for different values of our scale $M_{\chi}$. 
We notice that values of $\sin^{2}{\theta}_{atm}$ close to the mean $\sim 0.5$ are harder to find than the extremes in this model. 
This is because the Higgs mass constraint disfavors points in the parameter 
space where $\lambda_{1}=\lambda_{2}$, for which we have $\sin^{2}{\theta}_{atm}=\frac{1}{2}$. 
This can be seen on the right side of fig. \ref{sin2solatm}. Notice that our scan always respects neutrino experimental values, 
where the criteria used is that each of the 6 observables lies within its 3$\sigma$ experimental range, and then we compute the normalized $\chi^{2}$ function with respect to the best fit, given the experimental results in  \cite{Tortola:2012te,Forero:2014bxa}.
%
%
We show on fig. \ref{sin2reac_sol2atm2} the dependence of the solar and reactor angle against $\tilde{m}$ for different values of our scale $M_{\chi}$. We notice that small values of $\sin^{2}{\theta}_{reac}$ are favored, as expected in BRpV, while there is a less clear dependence for $\sin^{2}{\theta}_{sol}$.
%
%
We also notice the heavy dependence between the solar and atmospheric angles in this model on fig. \ref{sin2atmsin2sol}, quite independently of the value of $M_{\chi}$ and without fixing any of the other parameters of the model. 
We conclude that SS-BRpV can still deliver a good agreement with all experimental bounds.
%
%

\section{Lightest Neutralino Decay}

Another striking issue in supersymmetric models with violation of R-Parity is the instability 
of the lightest neutralino, which means that it is not a good candidate for Dark Matter. In these models, 
the gravitino is also unstable, but with a long lifetime and potentially it is a good candidate for 
Dark Matter \cite{Cottin:2014cca,Diaz:2011pc}. On the other hand, since the bilinear violation 
of R-parity provides an explanation for neutrino masses, the smallness of these ones implies 
that the RpV couplings must be relatively small, and hence, the BRpV decay rates are small. 
In this footing, Neutralino decay rates are related to the $\lambda_i$ parameters, which in turn 
relate these decays with the neutrino observables. If supersymmetry is realized by Nature and the lightest 
neutralino is observed, a precise measurement of its decay modes will be required. In the 
following, we study the branching ratios for the processes,
\begin{equation}
\chi^0_1\longrightarrow Z\nu_\ell \,,\qquad
\chi^0_1\longrightarrow W\ell\,,\qquad
\chi^0_1\longrightarrow h\nu_\ell
\end{equation}
where $\ell$ is any of the three leptons. 

Neutralino decays via sfermions are suppressed by the large sfermion masses. In Split Supersymmetry 
squarks and sleptons are heavy, and they are assumed to have a mass of the order $\widetilde m$. 
This scale is large, and the best approach in order to avoid large logarithms is to decouple heavy particles 
at that scale. Charginos and neutralinos are not necessarily as heavy, 
and in our approach we decouple them at a common scale $M_\chi\ll\widetilde m$. 
However, individual charginos and neutralinos, in practice, may have 
a mass different than $M_\chi$ depending on the actual values of the 
parameters $M_1$, $M_2$, and $\mu$.

In order to study the allowed parameter space for the lightest neutralino decay,
we perform a general scan, by varying the free parameters of the model as indicated in 
table \ref{tab1}.
\begin{table}
\begin{center}
\begin{tabular*}{0.6\textwidth}{@{\extracolsep{\fill}} c c c c }
\hline\hline
Parameter      & Minimum  &  Maximum  & Units \\ \hline
$\tan\beta$    & $1$      & $50$      & -     \\
$M_{\chi}$     & $200$    & $1100$    & GeV   \\
$\widetilde m$ & $10^{4}$ & $10^{10}$ & GeV   \\
$\lambda_i$    & $-0.1$   & $0.1$     & GeV   \\
$\mu_g$        & $0.002$  & $0.004$   & eV    \\
\hline\hline
\end{tabular*}
\caption{Scanned ranges for SS-BRpV parameters. 
\label{tab1}}
\end{center}
\end{table}

\begin{figure}[H]
\centering
\includegraphics[width=1.0\textwidth,angle=0]{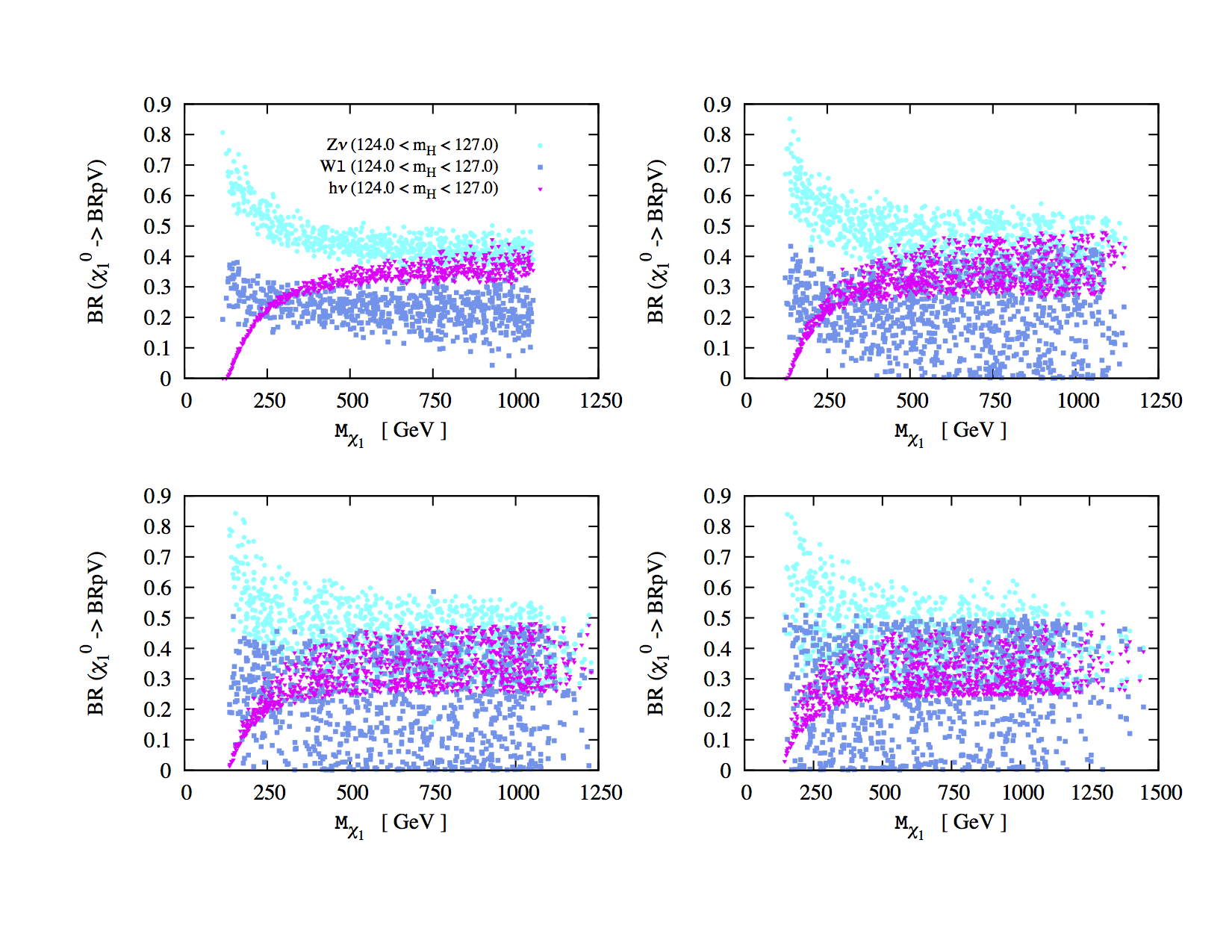}
\caption{ Branching ratios for the lightest neutralino as a function of the neutralino mass, with a decoupling scale
$M_\chi$. Neutralino/chargino parameters $M_1$, $M_2$, and $\mu$ are allowed to vary within $5\%$ (top left), $15\%$ (top right), $25\%$ (bottom left), and $50\%$ (bottom right)
of $M_\chi$, as explained in the text.
}
\label{BRDecoupling}
\end{figure}

In addition, we allow $M_1$, $M_2$, and $\mu$ to randomly vary above the 
decoupling scale $M_\chi$ according to the rule $p=M_\chi+rW$, where $p$ is one of these parameters, $r$ is a random 
number between 0 and 1, and $W$ is a window of variation that we take as $5\%$, $15\%$, $25\%$, or $50\%$
of $M_\chi$. For each of these windows, the lightest neutralino branching ratios are computed. 
In order to have consistency with the measured Higgs mass, we also impose a mass range for the SM Higgs of 
$124<m_h<127$ GeV. Furthermore, the best-fit values for neutrino oscillation physics, 
as given in the references \cite{Tortola:2012te,Forero:2014bxa} have been used in this calculation. 
Under these conditions, the results for the branching ratios are shown in fig.~\ref{BRDecoupling}.

%
%
From the figure, we clearly see the importance of the exact spectrum of neutral and charged fermions of the model. 
If the neutralinos and charginos have a mass close to a common scale $M_\chi$, then the 
lightest neutralino decay via neutral particles dominates. Nonetheless,
the situation is less clear if the spectrum is more spread out, and this is the case when we take
an increasing window $W$ where the parameters associated to these fermions can lie.
This result highlights the necessity to decouple the gauginos and higgsinos independently.

\begin{figure}[H]
\centering
\includegraphics[width=0.7\textwidth,angle=0]{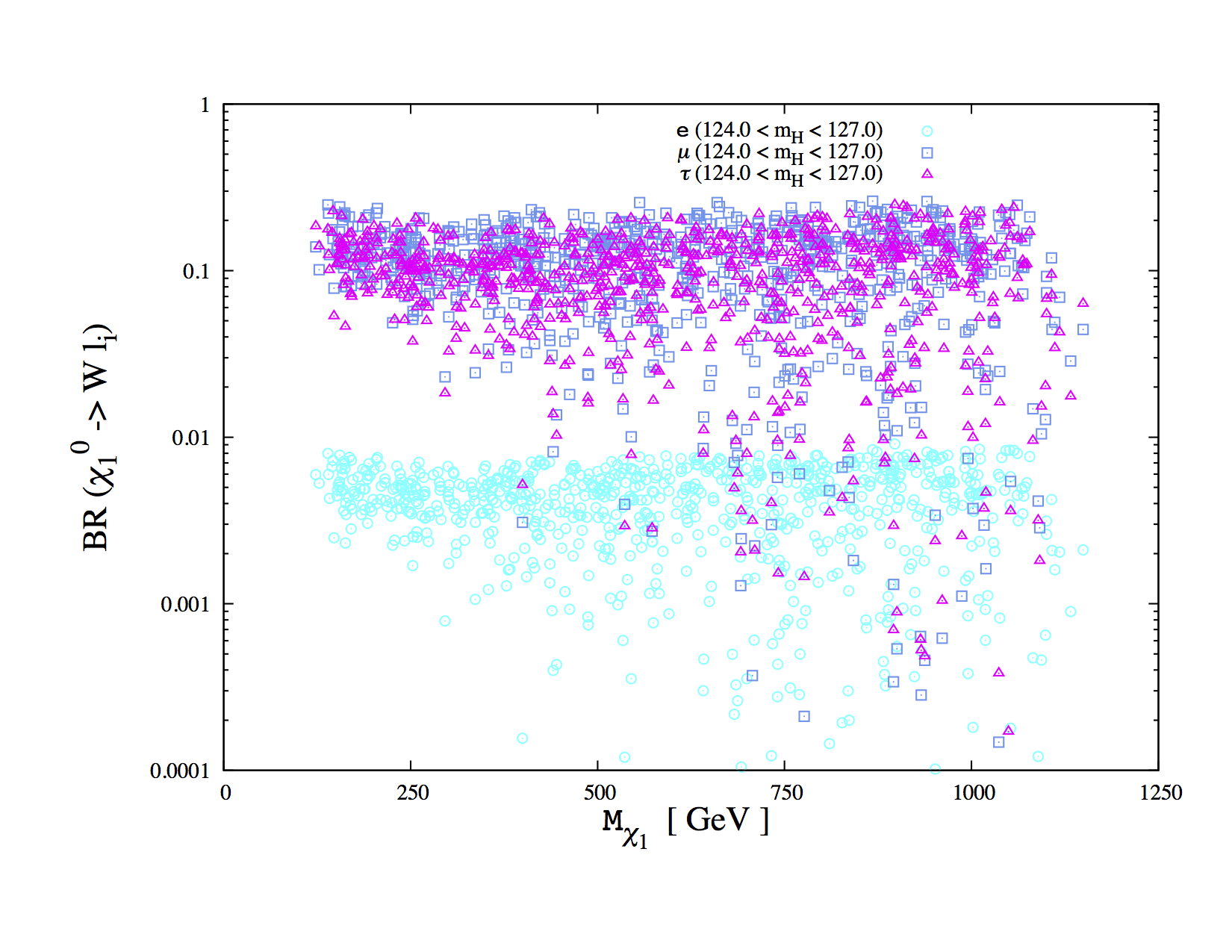}
\caption{ Branching ratios $BR(\chi^0_1\rightarrow W^\pm\ell^\mp)$ as a function of the neutralino mass for the scan
described in the text.
}
\label{BRWLScan}
\end{figure}

In turn, at fig.~\ref{BRDecoupling}, we see that the channel to charged particles may become relevant when compared to the 
other channels insofar the $W$ window is enlarged. In order to study this channel, in fig.~\ref{BRWLScan} we adopt a spread of $15\%$ and the three branching ratios 
$\chi^0_1\rightarrow W^\pm\ell^\mp_i$ for $\ell=e$, $\mu$, and $\tau$ are shown as a function of the lightest neutralino mass. We see that, for each of the points, the branching ratio $\chi^0_1\rightarrow W^\pm e^\mp$ is suppressed. This indicates that the term quadratic on $\lambda$ dominates over the $\mu_g$ term, since a small $BR(\chi^0_1\rightarrow W^\pm e^\mp)$ is associated 
to a small value for the neutrino reactor angle $\theta_{13}$ \cite{Guo:2007ug}. The other two branching 
ratios can be as large as 20-30\%, and any of both can be the largest.

%
\begin{figure}[H]
\centering
\includegraphics[width=0.7\textwidth,angle=0]{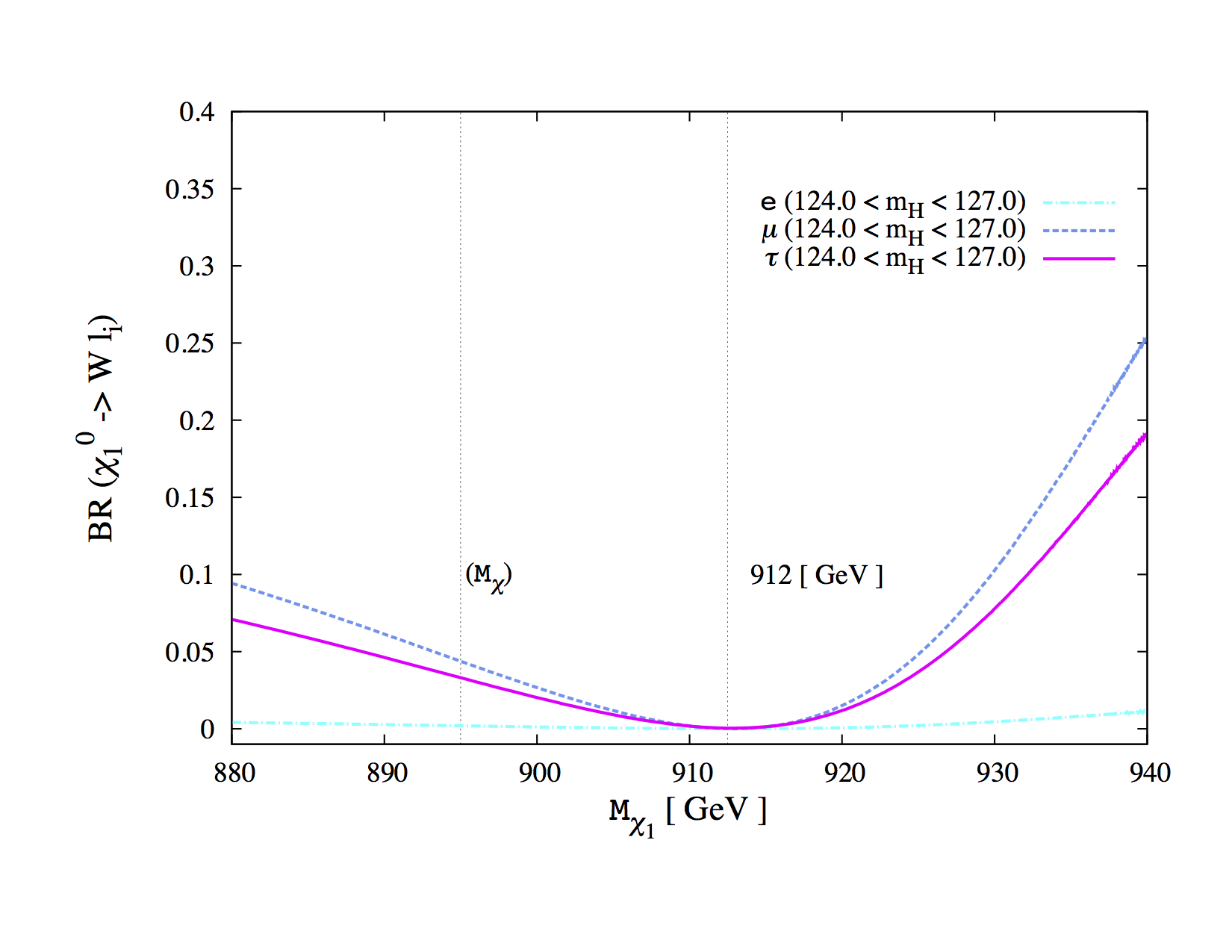}
\caption{ Branching ratios $BR(\chi^0_1\rightarrow W^\pm\ell^\mp)$ as a function of the lightest neutralino mass, 
with parameters fixed in SS-BRpV according to table \ref{tab2}. The vertical dotted lines 
show the value for the lightest neutralino mass where the curves hit zero, at $M_{\chi_1} \sim 912$ GeV,
and the value of the scale $M_\chi$.
}
\label{BRWLLines}
\end{figure}

%
In order to have a better appreciation of the dependency of the three branching ratios 
we are studying, in fig.~\ref{BRWLLines} we fix all the parameters with the exception of $M_1$, plotting 
the BR as a function of the neutralino mass. We choose to fix the parameters as it is indicated in table 
\ref{tab2} within a spread of 25\%. We see again that $BR(\chi^0_1\rightarrow W^\pm e^\mp)$ is suppressed, 
and for the chosen parameters $BR(\chi^0_1\rightarrow W^\pm\mu^\mp)>BR(\chi^0_1\rightarrow W^\pm\tau^\mp)$, 
which lie between $10\%$ and $25\%$ (passing through zero) for a neutralino mass varying between 880 and 940 GeV.
%
%
%
\begin{table}
\begin{center}
\begin{tabular*}{0.6\textwidth}{@{\extracolsep{\fill}} c c c }
\hline\hline
Parameter               & Value      & Units \\ \hline
$M_2$                   & $1010$      & GeV   \\
$M_3$                   & $1500$     & GeV   \\
$\mu$                   & $1020$     & GeV   \\
$\tan\beta$             & $2.21$      & -     \\
$M_{\chi}$              & $895.31$      & GeV   \\
$\log_{10}\widetilde m$ & $4.65$     & GeV   \\
$\lambda_1$             & $0.00110$  & GeV   \\
$\lambda_2$             & $0.00526$  & GeV   \\
$\lambda_3$             & $-0.00456$ & GeV   \\
$\mu_g$                 & $0.00304$  & eV    \\
\hline\hline
\end{tabular*}
\caption{Chosen parameters for fig.~\ref{BRWLLines}.
\label{tab2}}
\end{center}
\end{table}

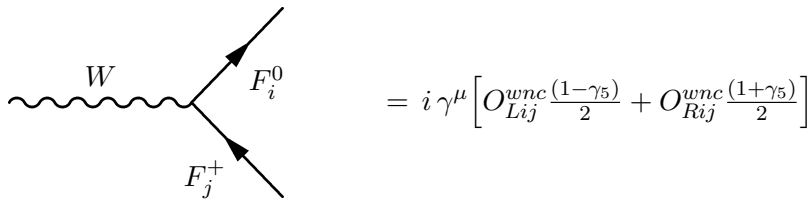
\begin{figure}[H]
\centering
\hspace{3cm}\begin{minipage}{0.45\linewidth}
\begin{fmffile}{photon} 
\begin{fmfgraph*}(40,25)
\fmfkeep{photon}
\fmfleft{i1}
\fmfright{o1,o2}
\fmf{photon,label=$W$}{w1,i1}
\fmf{fermion,label=$F^+_j$}{o1,w1}
\fmf{fermion,label=$F^0_i$}{w1,o2}
\end{fmfgraph*}
\end{fmffile}
\end{minipage}
\ \ 
\hfill
\hspace{-5cm}\begin{minipage}{0.45\linewidth}
$=\,i\,\gamma^\mu\Big[O^{wnc}_{Lij}\frac{(1-\gamma_5)}{2}+
O^{wnc}_{Rij}\frac{(1+\gamma_5)}{2}\Big]$
\end{minipage}
\caption{ Neutralino coupling to $W$ boson in SS-BRpV.}
\label{feynman}
\end{figure}

One particular feature on fig.~\ref{BRWLLines}, is that the three lepton 
branching ratios go to zero at the same value of $M_{\chi_1}$, in this case, around $912$ GeV. 
In order to understand this zero, we study the neutralino couplings to the
$W$ boson. The general coupling is given by fig.~\ref{feynman}.

\unitlength = 1mm

If we focus on $F^0_i\rightarrow\chi^0_i$ and $F^+_j\rightarrow\ell^+_j$ in the 
$WF^0F^+$ vertex, the couplings become $O^{wnc}_{Lij}\rightarrow O^{w\chi\ell}_{Lij}$ and
$O^{wnc}_{Rij}\rightarrow O^{w\chi\ell}_{Rij}$, with

\begin{eqnarray}
O^{w\chi\ell}_{Lij} &=& g \left[ N_{i2} \xi_L^{j1} + \frac{1}{\sqrt{2}} N_{i3} \left(
\xi_L^{j2} - \xi_{j3} \right) - \frac{1}{\sqrt{2}} \left( N_{i1} \xi_{j1}
+ N_{i2} \xi_{j2} + N_{i4} \xi_{j4} \right)
\right] \equiv \widetilde O^{w\chi\ell}_{Lj} \lambda_i
\nonumber\\
O^{w\chi\ell}_{Rij} &=& 0
\label{tildeOwchiell}
\end{eqnarray}

The quantities $N_{ij}$ are the components of the $4\times4$ matrix that diagonalizes the neutralino 
sector in the neutralino-neutrino mass matrix. The quantities $\xi_L^{ij}$ and $\xi_{ij}$, that
parametrize the chargino-charged lepton and neutralino-neutrino mixing respectively, can be found in 
ref.~\cite{Cottin:2014cca} and from their definition we see that the couplings in eq.~(\ref{tildeOwchiell}) 
are proportional to the parameters $\lambda_i$ defined below eq.~(\ref{detNeut}). If we make the following 
approximations: (i) motivated by the graph itself we assume the lightest neutralino is gaugino-like, (ii) 
we neglect the running of the $\tilde g$ parameters, and (iii) we assume that $v\ll M_1\sim M_2<\mu$, we obtain,
\begin{equation}
\widetilde O^{w\chi\ell}_{Lj} \approx \frac{gvc_\beta}{2\sqrt{2}\mu M_2}(g'N_{11}+gN_{12})
\end{equation}
We see that the coupling in eq.~(\ref{tildeOwchiell}) is proportional to $\lambda_i$ and to 
$\widetilde O^{w\chi\ell}_{Lj}$. In generating fig.~\ref{BRWLLines} we have kept $\lambda_i$ constant, 
thus the whole coupling of $W$ to charged fermions goes to zero at a point independent of the charged lepton
because the combination $(g'N_{11}+gN_{12})$ goes to zero. In other words, the neutralino does not couple to 
the $W$ gauge boson at this point. The fact that the neutralino decay mode to charged leptons may be 
suppressed in this model has implications on the choice of the decay mode in searches at the LHC.

\section{Summary}

We have studied the effect of a Higgs boson of mass $124<m_H<127$ GeV, motivated by measurements 
at the LHC, on a Split Supersymmetric model with Bilinear R-Parity Violation. We have checked
that the Higgs boson mass forces the split supersymmetric scale to be rather low, 
$\widetilde m < 10^{6}$ GeV, with a smaller influence from the gaugino mass.
Any value of $\tan\beta$ within $1<\tan\beta<50$ is allowed,
including the special case of $\tan\beta=1$, which holds possible as long as we give up 
gauge coupling unification, with extra new physics appearing at the scale ($>\widetilde m$) 
where the top Yukawa coupling becomes non-perturbative. 

We constrain neutrino parameters in this model, given the experimental results on neutrino
observables and Higgs mass. We find that independently of the 
allowed value of the mass squared differences, the Higgs mass grows with $\tilde{m}$ and rather small values of
$\tilde{m}$ are preferred, given the Higgs mass constrain. We also notice the effects of imposing a
Higgs mass constrain in the neutrino mixing angles. We find a striking 
dependence between the solar and atmospheric angles in this model, where strong constrains on 
$\sin^{2}{\theta}_{sol}$ limits $\sin^{2}{\theta}_{atm}$ quite independently of the chargino/neutralino decoupling scale. 
We still find points in the allowed parameter space 
in good agreement with all experimental bounds.

Finally, we have studied the two-body decays of the lightest neutralino in detail, in particular,
the effects of the exact spectrum of neutralinos and charginos and their decoupling. 
In general, the neutralino branching ratios are dominated by the channel to neutral particles, 
but insofar the exact spectrum is more spread around a larger decoupling scale, there may be other 
hierarchies for the neutralino branching ratios where the channel to charged particles may be more 
relevant. This issue indicates that the decoupling of charginos and neutralinos should be performed 
taking into account their exact spectrum. In addition, we conclude that future decaying 
neutralino searches at the LHC in this model should focus first on decays to neutral fermions. 

\section*{Acknowledgments}
{\small 
This work was partly funded by Conicyt-Fondecyt Regular grants 1100837 and 1141190. 
GC was funded by the postgraduate Conicyt-Chile Cambridge Scholarship 84130011.
NR was funded independently by Proyecto Anillo ACT1102, proyecto regular
Fondecyt 1141190, and by Becas Chile (Conicyt), Postdoctorado en el Extranjero 
(conv. 2014) num. 74150028. SO was funded by the postgraduate Conicyt Becas Chile.
}

\end{document}